\documentstyle[12pt]{article}
\newcommand{\gsim}{%
\mathrel{%
\setbox0=\hbox{$>$}
\raise0.6ex\copy0\kern-\wd0
\lower0.65ex\hbox{$\sim$}
}}
\begin{document}
\rightline{IPNO/TH 95-65}

\bigskip
\begin{center}
{\bf
HOW TO SEE THE CHIRAL STRUCTURE OF QCD
VACUUM IN LOW ENERGY $\pi-\pi$ SCATTERING
\footnote{Invited talk at the "Workshop on Physics and Detectors
for Da$\Phi$ne", Frascati 4-7 April 1995}
}

\bigskip
{
Jan STERN\\
{\em Division de Physique Th\' eorique\footnote{Unit\'e de Recherche
des Universit\'es Paris XI et Paris VI associ\'ee au CNRS}
, Institut de Physique
Nucl\' eaire,
91406 Orsay Cedex}\\
}
\end{center}
\baselineskip=14.5pt

\medskip
\begin{abstract}
Precise measurement of the $\pi\pi$-phase shift $\delta^0_0(E)$
 at very low energies would provide, for the first time, the
experimental evidence in favour of or against the existence of
a large quark condensate in the QCD vacuum, which is
standardly postulated as the mechanism of the spontaneous
breakdown of chiral symmetry in QCD. The contribution of
Da$\phi$ne to this discovery could be decisive. \end{abstract}

\baselineskip=17pt

\section{-- INTRODUCTION}

\par\indent
The origin and the dynamics of the spontaneous breakdown of
various symmetries in the Standard Model is not yet fully
understood. Different theoretical scenarios are conceivable
and crucial experimental tests are still missing. In the
electroweak sector, these tests require energies not reached
so far : The question is that of the existence of an
elementary Higgs field, of its self-interaction and of its
coupling to fermions. In the QCD sector, one meets a similar
situation concerning the spontaneous breakdown of chiral
symmetry (SBCHS). Here too, the standardly postulated
mechanism of symmetry breaking - the formation of a condensate
of quark-antiquark pairs in the QCD vacuum - has not so far
been tested experimentally. Such tests are possible within new
low-energy experiments requiring a precision not reached so
far. In this talk this point will be illustrated on the
example of low-energy $\pi\pi$-scattering which is
experimentally accessible in $K_{e4}$ decays. In this domain
Da$\phi$ne-Kloe can reach a decisive improvement and provide
the first experimental evidence in favour of or against the
existence of a strong quark condensation in the QCD vacuum.

\section{-- CHIRAL SYMMETRY IN QCD}
In the limit $m_u=m_d=m_s=0$ of massless light quarks, the QCD
 Lagrangian is invariant under the group of transformations
\begin{equation}
\psi\equiv
\pmatrix{u\cr
d\cr
s\cr}
\to
\Bigl\{
\frac{1}{2}(1-\gamma_5)g_L+\frac{1}{2}(1+\gamma_5)g_R
\Bigr\}e^{i\omega_v}\psi
\end{equation}
that consists of two independent $SU(3)$ rotations $g_L$ and
$g_R$ of the left-handed and right-handed components of quark
fields and of a common phase factor $e^{i\omega_v}$. This
global chiral symmetry  $SU(3)_L\times SU(3)_R\times U(1)_v$
implies the existence of 9 vector ($a=0,1\ldots 8$) and 8
axial ($i=1\ldots 8$) Noether currents
\begin{equation}
J^{a}_\mu(x)=\bar\psi\gamma_\mu\frac{\lambda^{a}}{2}\psi(x),
\quad
J^{i}_{\mu 5}(x)=\bar\psi\gamma_\mu\gamma_5
\frac{\lambda^{i}}{2}\psi(x)
\end{equation}
which are conserved
\begin{equation}
\partial^\mu J^{a}_\mu(x)=\partial^\mu J^{i}_{\mu 5}(x)=0\ .
\end{equation}
Provided the physical spectrum of the theory does not involve
massless quarks (confinement), the chiral symmetry
$SU(3)_L\times SU(3)_R\times U(1)_v$ must be spontaneously
broken down to the subgroup $U(3)_v$ generated by the 9 vector
currents $J^{a}_\mu$. This is known to be a mathematical
consequence of anomalous Ward identities$^{1)}$ and of the
vectorial character of the coupling of quarks and
gluons$^{2)}$. The statement of SBCHS has a double meaning :
First, the vacuum state $\vert 0>$ does not share all the
symmetries of the Lagrangian. In particular, the 8 axial
charges do not annihilate the vacuum :
\begin{equation}
\int
d^3\vec x\ J^{i}_{05}(\vec x,t)\vert 0>\not =0\ .
\label{vacu}\end{equation}
Next, there exist 8 massless
$J^p=0^-$ particles $\vert\pi^{i},\vec p>$ coupled to the 8
conserved axial currents :
\begin{equation}
<0\vert J^{i}_{\mu 5}\vert\pi^j,\vec p>=i\delta^{ij}
F_0p_\mu,\ p^2=0\ .
\end{equation}
The quantity $F_0$, which coincides with the chiral limit
$m_u=m_d=m_s=0$ of the pion decay constant $F_\pi=92.4$ MeV,
plays a fundamental role. It measures a long-range correlation
between the Noether currents in the vacuum of the massless
theory :
\begin{equation}
F_0^2\delta^{ij}=\frac{i}{3}
\int d^4x<0\vert T
\Bigl\{
J^{i}_{\mu 5}(x)J^{\mu,j}_5(0)-J^{i}_\mu(x)J^{\mu,j}(0)
\Bigr\}\vert 0>\ .
\label{corr}\end{equation}
The existence of this correlation implies the asymmetry of the
vacuum, c.f. Eq. (\ref{vacu}). Indeed, the operator in the
curly bracket on r.h.s. of Eq. (\ref{corr}) transforms as the
irreducible representation $\{8,8\}$ of the symmetry group
$SU(3)_L\times SU(3)_R$. Consequently, if the vacuum was
invariant, the matrix element (\ref{corr}) would necessarily
vanish. In other words, $F_0$ is an {\bf order parameter} :
its non zero value implies SBCHS. In fact, Goldstone theorem
guarantees that also the inverse statement is true : $F_0\not
=0$ is not only a sufficient but also a {\bf necessary}
condition of SBCHS.

The 8 Goldstone bosons are identified with the 8 lightest
$J^p=0^-$ particles $\pi,\ K,\ \bar K$ and $\eta$. The non zero
but small masses of pseudoscalar mesons reflect the
nonvanishing (running) masses of the light quarks u, d, s. The
mass term in the QCD Lagrangian
\begin{equation}
\L=\L_0-\bar\psi\ {\cal M}\ \psi,\quad {\cal M}=
\mbox{diag}(m_u,m_d,m_s)
\label{qcdl}\end{equation}
breaks the chiral symmetry explicitly. However, the chiral
symmetry still remains a good approximation and the mass term
in Eq.(\ref{qcdl}) can be treated as a small perturbation,
since, in the real world,
\begin{equation}
m_u,m_d,m_s\ll\Lambda_H\sim 1\mbox{GeV}
\end{equation}
where $\Lambda_H$ represents the mass scale of the first
massive bound states of the theory ($\rho,N\ldots$).

\section{-- THE QUARK CONDENSATE}

The theoretical facts summarized above represent unavoidable
consequences of the QCD Lagrangian. In spite of their
importance, these statements remain on a rather general level.
In particular, the basic fact of SBCHS, i.e. the non-zero
value of the correlator $F^2_0$ (\ref{corr}), does not by
itself imply a particular chiral structure in the QCD vacuum,
nor does it clarify the dynamical origin of SBCHS in QCD.
Further progress in this direction requires a consideration of
other order parameters. This point can be illustrated by an
analogy with the spin systems. In the latter case, the
spontaneous breakdown of rotation symmetry can be realized by
means of rather different types of magnetic order in the
ground state. {\bf Ferromagnets} are characterized by aligned
spins

\centerline{$\uparrow\quad\uparrow\quad\uparrow\quad\uparrow
\quad\uparrow\quad\uparrow\quad ,$}
whereas {\bf antiferromagnets} exhibit a rather different
magnetic structure of the type

\centerline{$\uparrow\quad\downarrow\quad\uparrow\quad
\downarrow\quad\uparrow\quad\downarrow\quad .$}

The order parameter which makes the distinction between these
two extreme cases is the average spontaneous magnetization
$<\vec m>$ : For ferromagnets $<\vec m>\not = 0$ and this
parameter plays a crucial role in the description of the
response of the system to an external magnetic field. On the
other hand, the magnetization of an antiferromagnet is
marginal or even vanishing to the extent that the ground state
approaches the N\'eel-type magnetic order. Hence, the
spontaneous magnetization is an example of order parameter
whose non-zero value is not necessary for the spontaneous
breakdown of symmetry to occur. Rather, it describes the
nature of the order in the ground state, i.e. the structural
details of the dynamics of symmetry breaking.

In QCD, the simplest order parameter which plays a similar role
 as the spontaneous magnetization of spin systems is the vacuum
condensate  $<0\vert\bar qq\vert 0>$ ($q=u,d,s$) defined in
the chiral limit. (In this limit, the definition of the quark
condensate is free of ultraviolet ambiguities.) It is
convenient to describe quark condensation by a parameter $B_0$
\begin{equation}
B_0=-\frac{1}{F^2_0}<0\vert\bar uu\vert0>=
-\frac{1}{F^2_0}<0\vert\bar dd\vert0>=
-\frac{1}{F^2_0}<0\vert\bar ss\vert0>\ ,
\label{cobo}\end{equation}
which has the dimension of mass. Since the operator $\bar qq$
carries anomalous dimension, $B_0$ is actually a running
quantity depending on the renormalization scale $\mu$. In
order to get a renormalization group invariant quantity, $B_0$
has to be multiplied by the quark mass. This fact explains,
why it is so difficult to detect quark condensation
experimentally : $B_0$ enters physical observables only
multiplied by a quark mass and consequently, it manifests
itself exclusively through tiny symmetry breaking effects.

One should, of course, understand within the framework of QCD
the interplay between SBCHS and quark condensation. In
particular, one should check and clarify the theoretical
possibility (suggested by the example of antiferromagnetism)
of SBCHS without formation of a quark condensate. The first
(rather modest) step in this direction can be reached,
expressing the order parameters $F_0^2$ (c.f. Eq. \ref{corr})
and $B_0$ (Eq. \ref{cobo}) in terms of the Euclidean
functional integral of QCD in a four-dimensional box of size
L. Integrating over quarks first (giving them a small mass m)
one finds, that for large volume and small masses, the order
parameters of SBCHS $F_0^2$ and $B_0$ are both determined by
the small eigenvalues of the Dirac operator
\begin{equation}
H(A)\phi_n=\lambda_n(A)\phi_n,\quad
H(A)=\gamma_\mu(i\partial_\mu+A^{a}_\mu t^{a})
\label{diop}\end{equation}
averaged over all gluonic
configurations $A_\mu^{a}(x).^{2)}$. (Notice that in Euclidean
space, $H^+=H$. Since $\{\gamma_5,H\}=0$, the spectrum
$\lambda_n$ is symmetric around 0.) In particular, the crucial
question is how dense does the spectrum become at the infrared
end as $L\to \infty^{3)}$. Three types of behavior appear as
particularly interesting. They are all characterized by
different thermodynamical properties (different dependence on
L and on m) and, in this sense, they appear as possible
distinct phases.

\noindent\begin{tabular}{lp{11cm}}
\bf I. $\lambda(L)\sim{1\over L}$ &In this case, both $F_0$
and the quark condensate vanish. Chiral symmetry is not
spontaneously broken (i.e. quarks cannot be confined.) \\
\bf II. $\lambda(L)\sim 1/L^2F_0$ &In this case$^{5)}$ one can
have $F_0\not = 0$, but the quark condensate still vanishes as
$L^{-2}$ in the limit $L\to\infty,\ m\to 0$. Goldstone bosons
are formed, SBCHS takes place but quarks do not condense. \\
\bf III. $\lambda(L)\sim 1/L^4\psi$ &This is the minimal level
density that is needed for quarks to start condense$^{4)}$.
Both $F_0$ and $B_0$ are different from zero : $<\bar qq>\sim
-\psi$.\\
\end{tabular}
\vglue 0.5cm
This discussion does not tell us what actually does happen in
the QCD vacuum. However, it collects theoretical possibilities
and it illustrates how SBCHS (i.e. $F_0\not =0$) without quark
condensation can naturally arise within QCD. In fact, it is
conceivable that in practice, the "phases" II. and III.
coexist and compete, giving rise to a marginal quark
condensate of the order $B_0\sim 100$ MeV, or so.

It is standardly postulated that in QCD, SBCHS is triggered by
a strong quark condensation$^{6)}$ with the parameter $B_0$ of
the order or even larger than the bound-state scale
$\Lambda_H\sim 1$ GeV. (A typical value at $\mu=1$ GeV would
be $B_0\sim1.6$ GeV$^{7)}$. This believe is usually motivated
by the Nambu, Jona-Lasinio model$^{8)}$ in which, indeed,
SBCHS and quark condensation can hardly be dissociated.
However, the chiral structure of the NJL model differs from
that of QCD : In the NJL model, the mechanism of SBCHS is not
related to the spectral properties of the Dirac operator
(\ref{diop}). It is conceivable that the previous discussion
and the emergence of the "phase II" as a theoretical
possibility are in fact limited to vector - like theories such
as QCD.

The standard scenario of a large quark condensate seems to be
supported by existing lattice simulations performed at finite m
and L and subsequently, extrapolated to the limit $m\to 0,\
L\to\infty$ following extrapolation formulae valid exclusively
in the phase III. Lattice regularization is known to mistreat
chiral symmetry in one way or another, especially in the
quenched approximation in which the most significant results
are obtained. (For a recent review see Ref. 9)). For these
reasons, the lattice results concerning the size of $<\bar
qq>$ deserve an independent experimental test.

The question of the strength of quark condensation has to be
settled experimentally. At present, none of the characteristic
consequences of the standard strong condensation scenario has
been experimentally tested. Actually, there is no single
experimental fact available that would allow to eliminate the
alternative possibility II of SBCHS without quark condensation.

\section{-- QUARK MASSES}

The amount of quark condensation has immediate consequences
for the relationship between pseudoscalar meson and (current)
quark masses. For $m_q\ll\Lambda_H$, one has for instance
\begin{equation}
\begin{array}{ll}
M^2_{\pi^+}=(m_u+m_d)B_0+(m_u+m_d)^2A_0+\ldots\\
M^2_{K^+}=(m_u+m_s)B_0+(m_u+m_s)^2A_0+\ldots\\
\end{array}
\label{maqu}\end{equation}
where the dots stand for chiral log's (irrelevant for the
present, merely qualitative discussion) and for higher order
terms. A similar expression holds for the $\eta$-mass, except
that at the quadratic level, a new constant appears reflecting
the axial U(1) anomaly and the $\eta-\eta'$ mixing. $A_0$ is
yet another order parameter of SBCHS : $F^2_0A_0$ represents a
long range correlation between scalar and pseudoscalar quark
densities of the type similar to (\ref{corr}). $A_0$ can be
roughly estimated using sum rules : $A_0\sim 1\div5$.

In view of the previous discussion, let us distinguish three
cases :  i) The (standard) case of a {\bf large condensate}
characterized by the condition
\begin{equation}
m_q\ll m_0=\frac{B_0}{2A_0}\sim\Lambda_H
\label{cond}\end{equation}
i.e. by the dominance of the first term in the expansion
(\ref{maqu}). ii) The case of {\bf vanishing condensate},
$B_0=0$ and finally, iii) the "mixed case" of a {\bf marginal
condensate} defined by
\begin{equation}
m_q\sim m_0=\frac{B_0}{2A_0}\ll\Lambda_H\ .
\end{equation}
In the latter case the first and second terms in Eq.
(\ref{maqu}) may become of a comparable size.

\subsection{\underline{The quark mass ratio $r=\frac{m_s}
{\widehat m},\widehat m=\frac{1}{2}(m_u+m_d)$}}

In the large condensate case this ratio is
predicted$^{10), 12)}$ to be
\begin{equation}
r=r_2=2\frac{M^2_K}{M^2_\pi}-1+\mbox{small corrections}
\simeq 26\ .
\label{rdeu}\end{equation}
If the condensate vanishes, Eq. (\ref{maqu}) implies
\begin{equation}
r=r_1=2\frac{M_K}{M_\pi}-1+\mbox{small corrections}
\simeq 6.3\ .
\label{rrun}\end{equation}
In the case of a marginal condensate, the ratio $r$ can take
 any value between the two extremes $r_1$ and $r_2$.
\begin{equation}
r_1\le r\le r_2\ .
\end{equation}

\subsection{\underline{The Gell-Mann, Oakes, Renner ratio
$x=\frac{(m_u+m_d)B_0}{M^2_\pi}$.}}

In the large condensate alternative, x is predicted to be close
 to 1$^{6),11)}$:
\begin{equation}
x=1+\frac{1}{32\pi^2}\frac{M^2_\pi}{F^2_\pi}\bar l_3+\cdots\ ,
\label{coal}\end{equation}
where $\bar l_3$ is one of the $SU(2)\times SU(2)$ low energy
constants introduced by Gasser and Leutwyler$^{11)}$. It is
expected to be of the order of unity. Comparing with Eq.
(\ref{maqu}), one finds
\begin{equation}
\bar l_3=-32\pi^2F^2_\pi\frac{A_0}{B^2_0}\ .
\label{lbar}\end{equation}
Hence, for $B_0\to 0,\ \bar l_3$ is naturally expected to be
large and negative ($A_0>0$). In this case the expansion
(\ref{coal}) breaks down, since it treats $m_q/m_0$ as a small
quantity. In the case of a marginal condensate, the GOR ratio
can actually take any value between 0 ($B_0=0$) and 1,
depending on the quark mass ratio $r$ :
\begin{equation}
x=\frac{(r-r_1)(r+r_1+2)}{r^2-1}+\mbox{small corrections}\ .
\label{rati}\end{equation}

\subsection{\underline{The $\eta$-mass}}

In the case of a large condensate, the $\eta$-mass is related
to the $\pi$ and $K$-masses by the well-known Gell-Mann Okubo
formula, modulo higher order corrections (including the
$\eta-\eta'$ mixing) which are hard to estimate in advance.
For smaller $B_0$ this relationship is lost. The reason is
that in this case, the unknown anomaly and $\eta-\eta'$ mixing
contributions are, in principle, of a comparable size as the
quark condensate contribution.

\subsection{\underline{The running quark mass $\widehat
m(\mu)$}}

The standard source of information on the magnitude of running
quark masses are the QCD sum rules (see e.g. $^{7),14)}$). A
model independent evaluation of the sum rule for $\widehat m$
is at present problematic, due to the complete absence of
experimental informations on the size and shape of the
spectral function associated with the divergence of the axial
current $\bar u\gamma_\mu\gamma_5d$, beyond the one-pion
contribution. Models for this spectral function that are {\bf
based on the large condensate hypothesis} lead to the
value$^{7)}$ $\widehat m$ (1 GeV) = $(6\pm 2)$ MeV. In the
low-condensate alternative, the spectral function is expected
to be considerably larger$^{13)}$ leading to a value of
$\widehat m$ $3\div 4$ times the value given above. On the
other hand, the sum rule determination of $m_s-\widehat m$
involves the divergence of the vector current $\bar
s\gamma_\mu u$ for which more experimental informations are
available. The resulting value$^{14)}$ $m_s-\widehat m=(184\pm
32)$ MeV (at $\mu=1$ GeV) is likely to be rather independent
of the strength of quark condensation. The question of the
size of $\widehat m$ thus becomes closely related to the
question of the magnitude of the ratio $r=m_s/\widehat m$ :
\begin{equation}
\widehat m\mbox{ (1 GeV)}\simeq \frac{184\pm
32}{r-1}\mbox{ (MeV)}\ .
\end{equation}

\section{-- CHIRAL PERTURBATION THEORY}

We have seen that different alternatives of quark condensation
manifest themselves through different values of the quark mass
ratio $r=m_s/\widehat m$, the GOR ratio $x=\frac{2\widehat m
B_0}{M^2_\pi}$ and, last but not least, the running quark mass
$\widehat m(\mu)$. Up to higher order corrections, the ratio
$r$ can be, for instance, measured comparing the observed
deviations form the Goldberger-Treimann relation in 3
different channels$^{15)}$. The current data suggest a value
of $r$ smaller by at least a factor 2 than the standard
$r\simeq 26$, but the uncertainties of this "determination"
are large. Similarly, the issue of the magnitude of $\widehat
m$ can, in principle, be settled, measuring with a high degree
of accuracy the tiny azimuthal asymmetries in the decay
$\tau\to 3\pi+\nu_\tau^{13)}$ - a project requiring a
Tau-Charm-Factory or a similar device. In addition to such
projects, the experimental determination of the strength of
quark condensation requires a systematic model independent
parametrization of various low-energy observables in terms of
the quark mass ratio $r$ and/or the GOR ratio x. Such a
prametrization is provided by the (generalized) Chiral
Perturbation Theory.

Chiral perturbation theory (CHPT) is a systematic low-energy
expansion of QCD correlation functions in powers of external
momenta p and quark masses $m_q$ such that the ratio $p/M_\pi$
is kept of the order of 1. It is based on the Ward identities
of the therory, SBCHS and on generalities such as analyticity,
crossing symmetry and unitarity. It is completely model
independent. The unknown features of the low-energy QCD
dynamics are parametrized by a set of low-energy constants
such as $F_0,\ \widehat mB_0,\ m_sB_0,$ etc.

	At low energies the only relevant degrees of freedom are
those of Goldstone bosons. This fact has allowed a
reformulation of CHPT as a low-energy effective theory (LEET)
based on the most general effective Lagrangian compatible with
the chiral symmetry$^{16)}$. The use of this technical device
has greatly simplified the practice of CHPT and it is at the
origin of its rapid development during the last 12
years$^{11), 12), 17)}$. LEET is still completely model
independent and it is equivalent to QCD, provided the
low-energy constants of $\L_{eff}$ are properly associated
with the order parameters of SBCHS in QCD. However, the
identification of all terms in $\L_{eff}$ contributing to a
given chiral order $0(p^d)$ depends on the strength of quark
condensation.
\eject

The standard version of CHPT$^{11), 12)}$ is based on the
hypothesis of a large quark condensate, as defined by the
condition (\ref{cond}) : symmetry breaking effects are
expanded in powers of both $m_q/\Lambda_H$ {\bf and}
$m_q/m_0$. In the case of a marginal or vanishing condensate,
the condition (\ref{cond}) breaks down and the expansion of
the effective Lagrangian has to be reformulated in a way which
does not treat $m_q/m_0$ as a small quantity. This leads to
the generalized chiral perturbation theory$^{18)}$ (GCHPT),
which at each chiral order $0(p^d)$ includes more terms of
$L_{eff}$ than its standard counterpart. GCHPT constitutes the
proper theoretical framework for an unbiased experimental
determination of $r=m_s/\widehat m$ and/or of the GOR ratio
$x=\frac{2\widehat mB_0}{M^2_\pi}$.

\section{-- LOW ENERGY $\pi\pi$ SCATTERING}

The chiral expansion of the $\pi\pi$ amplitude $A(s\vert tu)$
starts at the order $0(p^2)$.

{\bf The first step} in this expansion goes back to the 1966
work of Weinberg$^{19)}$:
\begin{equation}
A(s\vert tu)=\frac{1}{F^2_\pi}(s-2\widehat mB_0)+0(p^4)\ .
\label{Wein}\end{equation}
The $\pi\pi$ scattering is the priviledged place to
investigate quark condensation precisely because the amplitude
A explicitly depends on $B_0$ already at the leading
order$^{18)}$. In the large condensate alternative, $2\widehat
mB_0$ in Eq. (\ref{Wein}) is replaced by $M^2_\pi$ (yielding
the original Weinberg's formula). In this case Eq.
(\ref{Wein}) leads to the $I=0$ s-wave scattering length
$a^0_0=0.16$ whereas in the alternative of vanishing
condensate, one would have  $a^0_0=0.27$. Let us recall that
the current experimental value is$^{20)}$
\begin{equation}
a^0_0\vert_{exp}=0.26\pm 0.05\ .
\end{equation}
\indent {\bf The second step} was performed in 1983 by Gasser
and Leutwyler$^{11), 22)}$ who have computed the $0(p^4)$
corrections to A {\bf within the framework of the large
condensate hypothesis}. At order $0(p^4)$, the amplitude A
depends on 4 low-energy constants $\bar l_1,\bar l_2, \bar
l_3,\bar l_4$. Three of them $(\bar l_1,\bar l_2,\bar l_4)$
can be determined from other sources (d-waves, $K_{e4}$
form-factors, scalar radius of the pion, $F_K/F_\pi\cdots$).
The constant $\bar l_3$ measures the deviation of the GOR
ratio x from 1, see Eq. (\ref{lbar}). Assuming the latter not
to exceed a few \%, $(\bar l_3=2.9\pm2.4)^{11)}$ the
scattering length increases from the Weinberg's value 0.16
towards $a^0_0=0.20\pm 0.01$. In order to reach the central
experimental value $a^0_0=0.26$, $\bar l_3$ would have to be
as large and negative as $\bar l_3\sim-70$. Notice that within
the low-condensate altenative, a large negative value of $\bar
l_3$ is naturally expected, due to Eq. (\ref{rati}).

{\bf The third step} was completed only very recently$^{23)}$
by Knecht, Moussallam, Fuchs and Stern. The amplitude A has
been calculated up to and including order $0(p^6)$ (i.e. to
the two-loop accuracy) {\bf independently of any prejudice
about the strength of quark condensation}. At this order, A
depends on 6 parameters $\alpha,\beta,\lambda_1,\lambda_2,
\lambda_3,\lambda_4$. It is of the form
\begin{equation}
\begin{array}{ll}
A(s\vert t,u)&=\alpha\frac{M^2_\pi}{3F^2_\pi}+\frac{\beta}
{F^2_\pi}(s-\frac{4}{3}M^2_\pi)+\\
&+\frac{1}{F^4_\pi}
\Bigl\{
\lambda_1(s-2M^2_\pi)^2+\lambda_2
[(t-2M^2_\pi)^2+(u-2M^2_\pi)^2]
\Bigr\}+\\
&+\frac{1}{F^6_\pi}
\Bigl\{
\lambda_3(s-2M^2_\pi)^3+\lambda_4
[(t-2M^2_\pi)^3+(u-2M^2_\pi)^3]
\Bigr\}+\\
&+K_{\alpha,\beta,\lambda_1,\lambda_2}(s\vert
t,u)+0[(\frac{p}{\Lambda_H})^8]\ ,
\end{array}
\label{sixp}\end{equation}
where K is a non-trivial two-loop function (displayed in
Ref. 23)), which depends non-linearly on the parameters
$\alpha,\beta,\lambda_1$ and $\lambda_2$. The $0(p^4)$
amplitude can be obtained from Eq. (\ref{sixp}), setting
$\lambda_3=\lambda_4=0$ and reducing the function K to its
one-loop counterpart$^{23)}$ : There is a one to one
correspondence between Gasser- Leutwyler's $0(p^4)$ constants
$\bar l_1,\bar l_2,\bar l_3,\bar l_4$ and the parameters
$\alpha,\beta,\lambda_1,\lambda_2$ to the $O(p^4)$ accuracy.

The parameters $\lambda_1,\lambda_2,\lambda_3,\lambda_4$ can
be determined from the sum rules, using the existing
$\pi\pi$-scattering data at energies $E>500$ MeV$^{23)}$ :
\begin{equation}
\begin{array}{lll}
\lambda_1=(-5.3\pm 2.5)\times 10^{-3}\ ,\quad &\lambda_2&=(+9.7\pm 1.0)\times
10^{-3}\ ,\\
\lambda_3=(+2.9\pm 0.9)\times 10^{-4}\ ,\quad &\lambda_4&=(-1.4\pm 0.2)\times
10^{-4}\ .\\
\end{array}
\label{quat}\end{equation}
The remaining two parameters $\alpha$ and $\beta$ encode the
information on the strength of quark condensation. Given
(\ref{quat}), $\alpha$ and $\beta$ have to be determined
experimentally. For instance, the fit to the existing $K_{l4}$
data$^{21)}$ gives
\begin{equation}
\alpha=2.16\pm 0.86\quad\beta=1.074\pm 0.053\ ,
\end{equation}
reflecting the large experimental error bars. (The data points
are shown on the Figure.) This fit corresponds to the value of
the scattering length $a^0_0=0.263\pm 0.052$ in a perfect
agreement with the previous determination$^{20)}$ from the
same data analyzed using Roy equations.

The parameter $\alpha$ strongly depends on the quark mass
ratio $r=\frac{m_s}{\widehat m}$
\begin{equation}
\alpha=1+6\frac{r_2-r}{r^2-1}+\delta\alpha\ ,
\end{equation}
whereas $\beta$ stays close to 1 for all $r$ :
\begin{equation}
\beta=1+\frac{1}{r-1}
\bigl(\frac{F^2_K}{F^2_\pi}-1\bigr)+\delta\beta\ .
\end{equation}
$\delta\alpha$ and $\delta\beta$ contain (small) Zweig rule
violating terms and the (small) higher order chiral corrections
- they are discussed in Ref. 23). In the large condensate
alternative, one has $r\simeq r_2$ (c.f. Eq. (\ref{rdeu}) and
$\alpha$ is close to 1. Standard version of CHPT gives
\begin{equation}
\alpha_{st}=1.04\pm 0.15\
,\quad\beta_{st}=1.08\pm 0.03\ .
\end{equation}
In the case of a marginal or vanishing condensate, $r$ gets
closer to $r_1\simeq 6.3$ (Eq. (\ref{rrun})) and $\alpha$ can
be as large as 4. Hence the problem is to distinguish
experimentally between the cases i) $1<\alpha\le 1.2$ (large
condensate) ii) $1.5\le\alpha\le 3$ (marginal condensate) and
iii) $\alpha\ \gsim\ 3.5$ which might be considered as a
signature of the pure phase II with no quark condensation.

Since now the expression of the amplitude is available to
orders $0(p^2),\ 0(p^4)$ and $0(p^6)$, one can check the
convergence of the chiral expansion. In the case of a large
condensate, the scattering length $a^0_0$ takes, for instance,
the values 0.16, 0.20 and 0.21 at order $0(p^2),\ O(p^4)$ and
$0(p^6)$ respectively. A similar convergence rate is observed
for other threshold parameters and for other values of
$\alpha$ and $\beta$. Actually, a rather good convergence
persists even above threshold, in particular, in the
low-energy range accessible in $K_{l4}$ decays. This is shown
in Fig. a, where the difference $\delta^0_0-\delta^1_1$ (for
$\alpha=2,\ \beta=1.08$) is drawn for three successive orders
$0(p^2),\ O(p^4)$ and $0(p^6)$. The overall good convergence
rate suggests that the two-loop formula (23) is enough
accurate to serve as a basis for the analysis of low-energy
$\pi\pi$ data.

{\bf The last step} will be hopefully performed in a near
future by experimentalists. The difference of phase shifts
$\delta^0_0(E)-\delta^1_1(E)$ is accessible in a model
independent way in a $K^+_{l4}$ experiment. In Fig. b are
plotted 3 predictions for  $\delta^0_0-\delta^1_1$
corresponding to $\alpha=1.04,\ \alpha=2$ and $\alpha=3$
respectively. The data points are those of the last 1977
$K^+_{l4}$ experiment$^{21)}$. The figure shows the accuracy
that is needed in order to clearly distinguish between these 3
cases. A preliminary study shows$^{24)}$ that this accuracy
could well be reached in the Da$\phi$ne-Kloe $K^+_{l4}$
experiment.

An interesting independent project exists at CERN$^{25)}$ :
It aims to determine the difference of scattering lengths
$\Delta=a^0_0-a^2_0$ to a 5\% accuracy, measuring the lifetime
of the $\pi^+\pi^-$ atom. Such independent information would
be extremely useful. For $\alpha=1.04,\ \alpha=2.16$ and
$\alpha=3$, $\Delta$ is predicted to take values
$\Delta=0.253,\ \Delta=0.290$ and $\Delta=0.327$ respectively,
with a typical error bar (not including uncertainty due to
electromagnetic corrections) $\pm 0.007$.

\newpage
\begin{center}
\vskip 0.5 truecm
{\bf Figure 1:}{\it The phase shift difference $\delta_0^0-\delta_1^1$
in the range of energies accessible in the $K_{l4}$ decays is shown
a) for increasing chiral orders, and  b) for several values of
$\alpha$ and $\beta$\hfill}
\end{center}
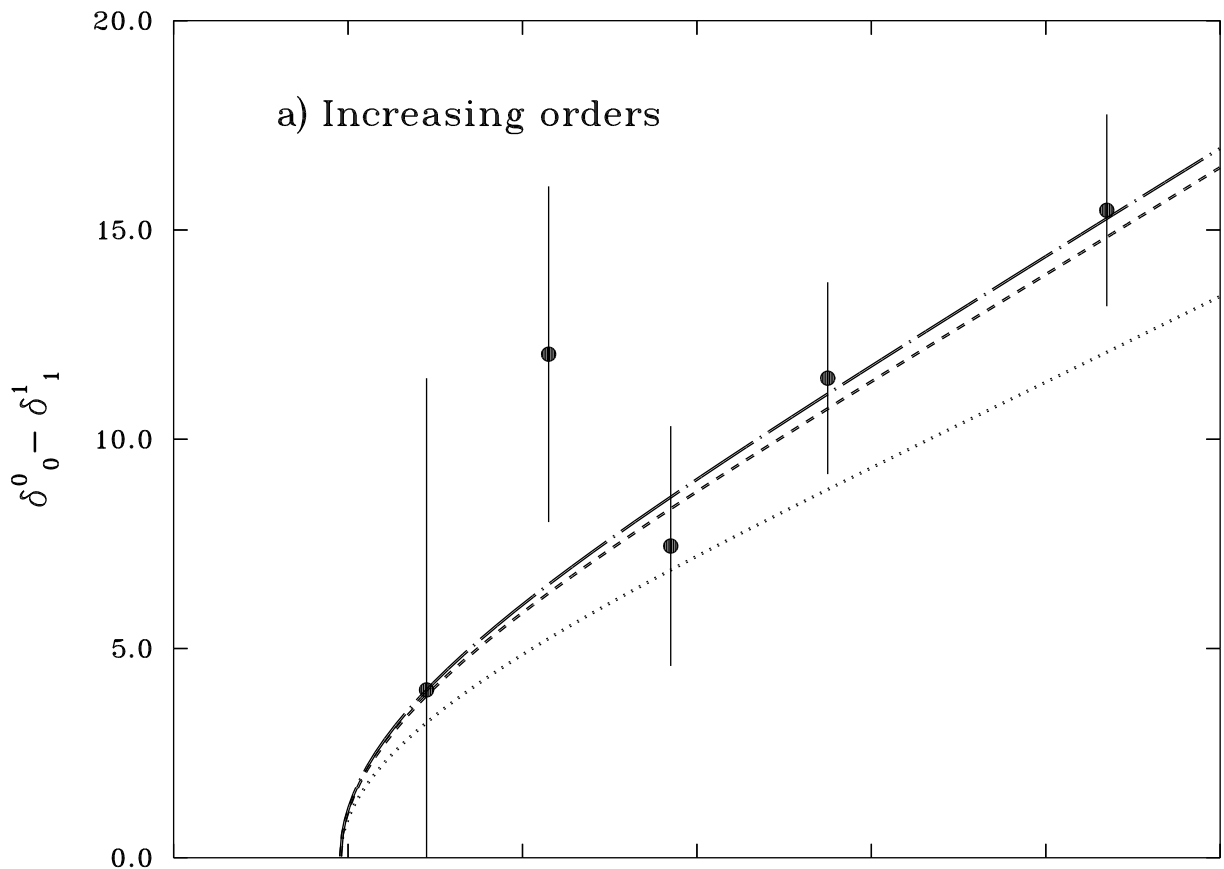
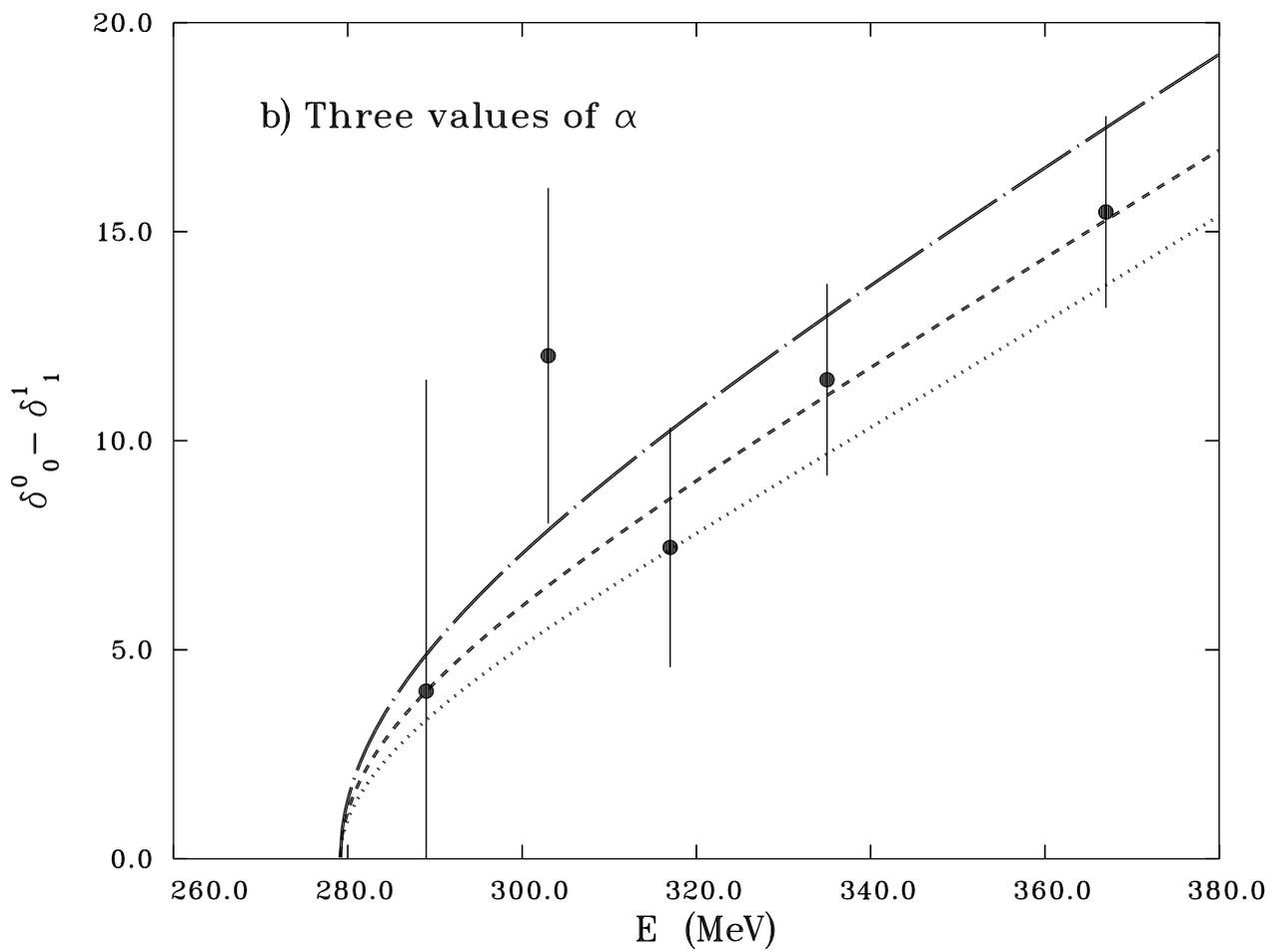
\end{document}